\documentclass[10pt]{article} 
\usepackage{pdfpages}
\usepackage{times}
\usepackage{etoolbox}

\usepackage{geometry}
\usepackage{fullpage}

\begin{document}

\title{
\vskip-10pt
Cultural Barriers to Software Productivity Practices at Los Alamos
}
\author{Charles R. Ferenbaugh
\thanks{
Mailing address:
Los Alamos National Laboratory, Mail Stop B287, Los Alamos, NM 87544, USA.
Email: {\tt cferenba@lanl.gov}.
The opinions expressed in this paper are solely those of the author.
}
\\
Applied Computer Science, CCS-7 \\
Los Alamos National Laboratory \\
}
\date{}
\maketitle
\thispagestyle{empty}

Since its beginnings during World War II, the nuclear weapons program
at Los Alamos National Laboratory has relied heavily on scientific
computation.
Large codes were written over many years, and significant efforts
were made to develop and implement cutting-edge physics methods.
For a long time, 
however, relatively little attention was given to
the computer science and software engineering aspects of the codes.
While some modern
software development practices had been adopted by the code teams
(version control, regression testing, automated builds and tests),
much more remained to be done.

In recent years, the code projects have given increased attention to
modern software productivity practices, due to a combination of factors:
\begin{itemize} \itemsep1pt \parskip0pt \parsep0pt
\item Increased size, complexity, and longevity of codes
\item Decreased funding and staffing levels
\item The transition from nuclear testing to predictive science simulations
\item The transition to new computer architectures such as
  Roadrunner~\cite{crf-se}
\end{itemize}
Responding to these changes,
existing code projects have taken steps to adopt some new practices,
while new and experimental projects have tried out other practices
as well.  As we did this, we found that some of the biggest barriers
to new practices were not technical but cultural.  Some aspects of
the institutional culture had evolved to fit traditional ways of doing
things, and worked against adoption of new practices.  Here are some of
the specific cultural issues we discovered.

\paragraph{Separated communities.}
Historically, physics code development and computer science%
\footnote{For brevity, I will lump together computer science, software
engineering, and other related disciplines under the label ``computer
science'' or ``CS'' in this paper.}
at LANL
have been done in isolation, by very different organizations with
different priorities.
This has led to misunderstandings between the two groups when they have
needed to interact.  Teams from the CS community have tried to introduce
changes to the code projects, and in some cases the changes proved to be
inappropriate.  Examples include
bleeding-edge language features (C++ expression templates, when these were
very new to the language standard and not well-supported by compilers)
and poorly-considered software processes (classic, process-heavy waterfall
development model, with no tailoring for a research environment).
These changes met with little success, and as a result, many on the CS side
saw the physicists as stuck-in-the-mud and resistant to change.
Meanwhile, the physicists saw the CS community as more interested
in experimenting with fancy new architectures and programming models
than in meeting mission requirements.
In both cases the reasons for the mistrust were perhaps exaggerated,
but not unjustified.

\paragraph{Management priorities.}
In the past, the main measure for success of code projects has been
``how much physics is in the code.''
Managers expected developers to implement physics features as quickly as
possible, and then move on to the next feature.
Managers and sponsors set
high-level, high-visibility milestones for the projects,
which nearly always consisted of
physics capabilities to be added or improved in the code.
In practice, developers knew what management's priorities were, and
when time and resources ran short (which was most of the time), they would
drop any work not needed for the physics task at hand.
Software quality practices were shortchanged or ignored (or never
even considered) in order to finish the physics.

\paragraph{Multiple commitments.}
Many modern, agile development approaches call for team members
to co-locate, to allow for practices such as
pair programming, group design reviews, and frequent
exchange of information.  Several of our recent efforts have attempted
to do this, with varying degrees of success.  
Most team members at LANL work part-time on multiple projects; when
many such people are on the same team, coordinating large blocks of
time for co-location becomes difficult.
Furthermore, many of the legacy projects have an
interrupt-driven culture, assuming that team members (especially key ones)
are available on demand.  This approach is understandable for projects
with a large code base and a small staff, often with only one or two
people who know key parts of the code.  Nevertheless, this made things
difficult for new projects whose members kept getting pulled away for
``emergencies'' on legacy projects.

\paragraph{Requirements versus implementations.}
Good software practice requires developers to capture product requirements
from customers and users, either formally or informally.
At LANL, the users have often developed
software themselves at some point in the past,
and they often think of requirements in terms of possible implementations.
Their not-quite-explicit thought process goes,
``I need capability~A.  I would implement~A using method~B.  Therefore,
B is a requirement.''  Sometimes~B just reflects the user's experience
(or lack thereof); sometimes there is sound reasoning behind the choice
to do~B, and developers really should adopt it or at least consider it.
In such a context, it is challenging for the developers to determine the
real requirements.

\paragraph{Experimental versus production code.}
Often physicists write their own software packages
to use as tools for simple data analysis, or to experiment
with new physics algorithms.  They use simple, ad-hoc software
processes which are sufficient for that context.
Meanwhile, projects that develop production codes require more detailed
processes, including such practices as good software design, documentation,
and testing.
When physicists work on such a project, there can be misunderstandings
caused by these two different (and legitimate) views of how much process
is needed to write software.
Furthermore, in research settings such as LANL,
production codes are often also used as testbeds for algorithmic
research.  This can mean that different levels of process are
appropriate for different tasks even within the same project and code base,
leading to some confusion if it's not clear which category applies
to a given task.

\paragraph{Verification and validation.}
The legacy codes at LANL have been developed and used over a period of
many years.  During that time, they have been used to run many tests,
and the results have been compared
to a variety of analytical solutions and experimental data.
This gives the code
a significant amount of verification and validation, both formally and
informally, increasing users' confidence in the outputs.
Meanwhile, developers from outside the physics community sometimes 
propose to quickly write new software to replace the legacy code,
and don't fully appreciate the amount of V\&V the legacy code has had.
When the physicists are reluctant to accept
the new code, this is sometimes mistaken for simple stubbornness,
and not understood as an issue of inadequate testing.
\\

With some time and experience, LANL has started to recognize
and work through some of the cultural issues identified above.
There is now more interaction between
the physics and CS communities, and this has led to greater
trust between the two groups and greater understanding of each other's
concerns.
New projects are started with a good balance between physics and CS team
members,
and legacy projects have brought new team members on board from the CS
side.  Several projects have changed their leadership structure to include
one physics lead and one CS lead, rather than
a single (physics) lead as in the past.
There is increased awareness at all levels of the management chain
that software issues require attention, especially where future
architectures are involved.

And, perhaps most importantly, we are exchanging ideas with other
software development communities, at LANL and elsewhere, that are dealing
with similar issues.
We look forward to more such interactions in the future.

\patchcmd{\thebibliography}{\section*{\refname}}{\subsection*{\refname}}{}{}

\end{document}